\documentstyle[12pt,a4,psfig]{article}
\begin{document}
\begin{sf}
\renewcommand {\thepage} { }
\renewcommand {\thefootnote} {\fnsymbol{footnote}}
\setcounter {page} {0}
\setcounter {footnote} {0}

\vspace*{1cm}
\vspace{5mm}
\begin{center}
\Large \bf
Photon tunneling \\[.5ex] through absorbing dielectric barriers
\end{center}
\vspace{5mm}
\begin{center}
Toralf Gruner and Dirk--Gunnar Welsch \\
Friedrich-Schiller-Universit\"at Jena,
Theoretisch-Physikalisches Institut \\
Max-Wien Platz 1, D-07743 Jena, Germany 
\end{center}
%
%\vspace{3cm}
\vspace{10mm}

\begin{center}
\bf Abstract
\end{center}
Using a recently developed formalism of quantization of
radiation in the presence of absorbing dielectric bodies,
the problem of photon tunneling through absorbing barriers is 
studied. The multilayer barriers are described in terms of 
multistep complex permittivities in the frequency domain which 
satisfy the Kramers--Kronig relations. From the resulting 
input--output relations it is shown that losses in the layers 
may considerably change the photon tunneling times observed in
two-photon interference experiments.
It is further shown that for sufficiently large numbers
of layers interference fringes are observed that cannot
be related to a single traversal time.
\vspace{0.5cm}

PACS number(s): 42.50.Ct, 73.40.Gk, 42.79.-e
\vspace{1cm}

\vfill

\newpage
\renewcommand {\thepage} {\arabic{page}}
\setcounter {page} {1}
\section{Introduction}
\label{intro}
Stimulated by recent experiments \cite{CKS,17b,1a}, the problem of 
photon tunneling through multilayer dielectric barriers has been of 
increasing interest. 
In order to answer the question of what is the 
time that is spent by a photon inside such a barrier, the effects of 
dispersion and absorption should be considered very carefully.
The calculations that have been performed so far are based on
real refractive indices of the layers \cite {CKS,17b,1a,4n},
so that a number of questions, such as the influence of absorption 
on the measured traversal times \cite{5n}, have been open.
It is well known that in frequency intervals where the dielectric 
layers are nearly transparent the action of each layer and
hence that of a multilayer barrier can be described in terms 
of unitary transformations that relate the operators of the 
outgoing fields to those of the incoming fields (see, e.g., 
Ref.~\cite{4}). These input--output relations and the underlying
quantization scheme (see, e.g., Refs.~\cite{1,2,3}) of course fail,
when the effects of absorption cannot be disregarded. 

Various approaches to the problem of quantization of 
radiation in the presence of absorbing dielectric bodies have
been developed \cite{8,8a,9,10,10a,11,11a,12,13,13a,14a,14}. 
In the present paper we use the quantization scheme
given in Refs.~\cite{14a,14}. It is based on a Green function
expansion of the operator of the (transverse) vector
potential and applies to radiation in both homogeneous and 
inhomogeneous dielectric matter. In this approach, the matter is 
described in terms of a complex permittivity (in the frequency 
domain), without using a particular microscopic model for the
matter. The only condition is that the permittivity satisfies
the Kramers--Kronig relations, because of causality. 
Applying the method to the calculation of input--output 
relations for radiation at absorbing multilayer dielectric barriers,
we can systematically study the effects of dispersion
and absorption on the propagation of single-photon wave packets
through such barriers.

The paper is organized as follows. In Sec.~\ref{Gre} 
the quantization scheme is outlined. In Sec.~\ref{Mul} the scheme is 
applied to radiation falling on multilayer dielectric barriers. 
The input--output relations derived are used in Sec.~\ref{Pt} in order 
to calculate barrier traversal times measurable in two-photon interference 
experiments. Finally, a summary is given in Sec.~\ref{Sum}. 

\section{Field quantization}
\label{Gre}
Let us consider linearly polarized light propagating in $x$\,direction
in an inhomogeneous linear dielectric medium. 
Introducing the (transverse) vector potential 
\begin{equation}
A(x,t) = \int_0^{\infty} {\rm d} \omega \,
e^{-i \omega t} A(x,\omega) + \mbox{c.c.}
\label{2.2a}
\end{equation}
($A$ $\!\equiv$ $\!A_y$),
the classical (phenomenological) Maxwell equations yield
\begin{equation}
\left[ \frac{\partial^2}{\partial x^2}
+ \frac{\omega^2}{c^2} \, \epsilon(x,\omega) \right] A(x,\omega) = 0,
\label{2.1}
\end{equation}
where 
\begin{equation}
\epsilon(x,\omega) = \epsilon_{\rm r}(x,\omega)
                        + i \, \epsilon_{\rm i}(x,\omega)
\label{2.7}
\end{equation}
is the complex permittivity which for 
inhomogeneous media, such as multilayer dielectric barriers, 
varies with $x$. Clearly, when the permittivity is complex, then 
Eq.~(\ref{2.1}) cannot be valid as an operator equation in quantum theory. 
On the other hand, it is well known that propagation of light in 
absorbing matter is unavoidably accompanied by noise. In a quantized 
version of Eq.~(\ref{2.1}) this noise source can be taken into account 
by introducing an operator noise current $ \,\hat{\!j}_{\rm n} $, 
so that \cite{14}
\begin{equation}
\left[ \frac{\partial^2}{\partial x^2}
  + \frac{\omega^2}{c^2} \, \epsilon(x,\omega) \right] \hat{A}(x,\omega) =
\,\hat{\!j}_{\rm n}(x,\omega),
\label{2.3}
\end{equation}
where
\begin{equation}
\hat{\!j}_{\rm n}(x,\omega) = \frac{\omega}{c^2}
\sqrt{\frac{\hbar}{\pi \epsilon_0 {\cal A}}\,\epsilon_{\rm i}(x,\omega)}
\,\hat{f}(x,\omega).
\label{2.4}
\end{equation}
Here, $\hat{f}(x,\omega)$ is a bosonic basic field, 
\begin{equation}
\left[
\hat{f}(x,\omega) , \hat{f}^{\dagger} (x',\omega')
\right] = \delta(x-x') \, \delta(\omega - \omega'),
\label{2.5}
\end{equation}
\begin{equation}
\left[
\hat{f}(x,\omega) , \hat{f} (x',\omega')
\right] = \left[
\hat{f}^{\dagger}(x,\omega) , \hat{f}^{\dagger} (x',\omega')
\right] = 0,
\label{2.6}
\end{equation}
and ${\cal A}$ is a normalization area in the $yz$\,plane.
Equation~(\ref{2.3}) is now an equation for the
the (Schr\"{o}dinger) operator $\hat{A}(x,\omega)$. The solution 
can be represented as
\begin{equation}
\hat{A}(x,\omega) = \int {\rm d} x'\, G(x,x',\omega) 
\,\hat{\!j}_{\rm n}(x',\omega),
\label{2.8}
\end{equation}
where $ G(x,x',\omega) $ is the classical Green function that
satisfies the equation
\begin{equation}
\left[ \frac{\partial^2}{\partial x^2}
+ \frac{\omega^2}{c^2} \epsilon(x,\omega) \right] G (x,x',\omega) =
\delta(x-x')
\label{2.9}
\end{equation}
and vanishes in the limit when $x\!\to\!\pm\infty$.
The quantization scheme ensures that the well-known
equal-time commutation relation
\begin{equation}
\left[
\hat{A}(x,t),\hat{E}(x',t)
\right] = - \frac{i \hbar}{{\cal A} \epsilon_0} \delta(x-x')
\label{2.10}
\end{equation}
is preserved \cite{14}.

\section{Input--output relations}
\label{Mul}
Let us consider a dielectric barrier consisting of $(N$ $\!-$ $\!2)$
layers ($N$ $\!\geq$ $\!3$), 
\begin{equation}
\epsilon(x,\omega) = \sum_{j=1}^N \lambda_j(x) \, \epsilon_j(\omega),
\label{3.1}
\end{equation}
where
$\epsilon_j(\omega)$ is the permittivity of the $j$th layer, and
\begin{equation}
\lambda_j(x) = \left\{
\begin{array}{ll}
1 &  {\rm if}\ x_{j-1} < x < x_j,  
\\
0 & {\rm otherwise}
\end{array}
\right.
\label{3.2}
\end{equation}
($x_0$ $\!\to$ $\!-$ $\!\infty$, $x_N$ $\!\to$ $\!\infty$). From
Eqs.~(\ref{2.2a}), (\ref{2.8}), and (\ref{2.9}), with $\epsilon(x,\omega)$
from Eq.~(\ref{3.1}), the operator $\hat{A}(x,\omega)$
can be rewritten as \cite{14}
\begin{eqnarray}
\lefteqn
{\hat{A} (x)= \sum\limits_{j=1}^N \lambda(x) \int_0^{\infty} {\rm d} \omega 
\,\sqrt{\frac{\hbar\beta_j(\omega) }
{4 \pi c \omega \epsilon_0 \epsilon_j (\omega) {\cal A}}} 
}
\nonumber
\\ & &
\hspace{6ex}
\times
\left[ e^{i\beta_j(\omega) \omega x/c} \, \hat{a}_{j+}(x,\omega) 
+ e^{-i\beta_j(\omega) \omega x/c} \, \hat{a}_{j-}(x,\omega) 
\right] + \mbox{H.c.},
\label{3.3}
\end{eqnarray}
where the quasi-mode operators $\hat{a}_{j\,+}(x,\omega)$ and 
$\hat{a}_{j\,-}(x,\omega)$,
which are associated with the (damped) waves propagating to the right 
and left, respectively, are related to the bosonic basic field as
\begin{eqnarray}
\lefteqn{
\hat{a}_{j \pm} (x,\omega) = 
\hat{a}_{j \pm} (x',\omega) \,
e^{\mp \gamma_j (\omega) \omega \left( x-x'\right)/c} 
}
\nonumber \\ && \hspace{5ex}
\pm \frac{1}{i} \sqrt{2 \gamma_j (\omega) 
\frac{\omega}{c}} 
\int_{x'}^x {\rm d}y \,
e^{\mp \gamma_j(\omega)\omega(x-y)/c}
e^{\mp i \beta_j (\omega) \omega y/c} \hat{f} (y,\omega)
\label{3.4}
\end{eqnarray}
($x_{j-1}$ $\!\leq$ $\!x$,$x'$ $\!\leq$ $\!x_j$). Here the notation
$\sqrt{\epsilon_j(\omega)}$ $\! =$ $\! n_j(\omega)$ $\!=$
$\beta_j(\omega)$ $\!+$ $\!i\,\gamma_j(\omega)$ is introduced.  

Using the commutation relation (\ref{2.5}), from Eqs.~(\ref{3.4})
we find that the quasi-mode operators $\hat{a}_{1+}(x,\omega)$,
$x$ $\!\leq$ $\!x_1$, and $\hat{a}_{N-}(x,\omega)$, $x$ $\!\geq$ $\!x_N$,
of the incoming radiation from the left and the right, respectively,
satisfy the commutation relations
\begin{equation}
\left[ \hat{a}_{1+} (x,\omega), \hat{a}^{\dagger}_{1+} (x',\omega') \right]
= \delta \left( \omega- \omega'  \right) 
\, e^{-\gamma_1(\omega) \omega \left| x - x' \right|/c},
\label{3.5} 
\end{equation}
\begin{equation} 
\left[ \hat{a}_{N-} (x,\omega), \hat{a}^{\dagger}_{N-} (x',\omega') \right]
= \delta \left( \omega- \omega'  \right) 
\, e^{-\gamma_N(\omega) \omega \left| x - x' \right|/c},
\label{3.6} 
\end{equation}
\begin{equation} 
\left[ \hat{a}_{1+} (x,\omega), \hat{a}^{\dagger}_{N-} (x',\omega') \right]
= 0.
\label{3.6a}
\end{equation}
Note that Eqs.~(\ref{3.5}) and (\ref{3.6}) agree with the commutation 
relations valid for the corresponding bulk materials. For vanishing absorption   
($\gamma_{1(N)} (\omega)$ $\! \to$ $\! 0$) the operators 
$\hat{a}_{1+}$ and $\hat{a}_{N-}$ 	
are independent of $x$ and ordinary free-field bosonic operators
\cite{14}.

The output operators
$\hat{a}_{1-}(x,\omega)$, $x$ $\!\leq$ $\!x_1$, 
and $\hat{a}_{N+}(x,\omega)$, $x$ $\!\geq$ $\!x_N$, 
can then be calculated step by step starting
from a single-slab plate ($N$ $\!=$ $\!3$). 
Using Eqs.~(\ref{3.3}) and 
(\ref{3.4}) and taking into consideration that
the vector potential must be continuously differentiable at
the interfaces, after some lengthy calculation we find that
the output operators can be expressed in terms of the input operators
and bosonic operator noise sources $\hat{g}_\pm(\omega)$ associated
with the losses in the barrier \cite{14a,19},
\begin{equation}
\left( \begin{array}{c}
\hat{a}_{1-}(x_1,\omega) \\ \hat{a}_{N+}(x_{N-1},\omega)
\end{array} \right)
= 
\tilde{\bf T}(\omega)
\left( \begin{array}{c}
\hat{a}_{1+}(x_1,\omega) \\ \hat{a}_{N-}(x_{N-1},\omega)
\end{array} \right)
+
\tilde{\bf A}(\omega)
\left( \begin{array}{c}
\hat{g}_{+}(\omega) \\ \hat{g}_{-}(\omega)
\end{array} \right)\!,
\label{3.10}
\end{equation}
\begin{equation}
\left[\hat{g}_\pm(\omega),\hat{g}^\dagger_\pm(\omega')\right] 
= \delta(\omega-\omega')
\label{3.10a}
\end{equation}
(the input and noise operators are commuting quantities).
Here the characteristic transformation matrix 
$ \tilde{\bf T}(\omega) $ describes the effects of transmission and 
reflection of the input fields \cite{BW}, whereas the losses inside 
the barrier give rise to an absorption matrix $\tilde{\bf A}(\omega)$. 
Explicit expressions for the matrices $ \tilde{\bf T}(\omega) $ 
and $\tilde{\bf A}(\omega)$
and the noise operators $\hat{g}_\pm(\omega)$ [as linear functionals of 
the field $\hat{f}(x,\omega)$ inside the barrier]
are given in Ref.~\cite{19}.

The input--output relations (\ref{3.10}) [together with Eq.~(\ref{3.4})]
lead to commutation relations for the output operators
$\hat{a}_{1-}(x,\omega)$, $x$ $\!\leq$ $\!x_1$, and 
$\hat{a}_{N+}(x,\omega)$, $x$ $\!\geq$ $\!x_N$, that differ, in general, 
from those in Eqs.~(\ref{3.5}) -- (\ref{3.6a}) for the input operators.
The difference decreases with increasing distances from the barrier.
In particular, it can be disregarded when the sourrounding medium can be 
regarded as beeing lossless and the input and output operators
are ordinary bosonic operators. 
In this case the relations
\begin{eqnarray}
\lefteqn{
\left| T_{11}(\omega) \right|^2 +\left| T_{12}(\omega) \right|^2
+ \left| A_{11}(\omega) \right|^2 + \left| A_{12}(\omega) \right|^2
}
\nonumber
\\ & &
\hspace{7ex}
= \left| T_{21}(\omega) \right|^2 +\left| T_{22}(\omega) \right|^2
+ \left| A_{21}(\omega) \right|^2 + \left| A_{22}(\omega) \right|^2
=1,
\label{3.11}
\\ & &
T_{11} (\omega) T^{\ast}_{21} (\omega)
+ T_{12} (\omega) T^{\ast}_{22} (\omega)
+ A_{11} (\omega) A^{\ast}_{21} (\omega)
+ A_{21} (\omega) A^{\ast}_{22} (\omega)
=0  \hspace{3ex}
\label{3.12}
\end{eqnarray}
can be shown to be valid, which ensure preservation of the
bosonic commutation relations.

\section{Photon tunneling}
\label{Pt}
To study the influence of dispersion and absorption on photon 
tunneling through multilayer dielectric barriers, let us
consider a two-photon experiment of the type described in
Ref.~\cite{CKS} (Fig.~1). Pairs of down-conversion photons
are directed by mirrors to impinge on the surface of a 
50\%:50\% beam splitter and the output coincidences are measured.
One photon ({\rm I}) of each pair travels through air, while the conjugate
photon ({\rm II}) passes a barrier. The coincidences attain a minimum when
the two photons' wavepackets overlap perfectly at the beam
splitter. This can be achieved by translating an appropriately
chosen prism in one arm of the interferometer in order to compensate
for the delay owing to the barrier.

Let us assume that the barrier is in the ground state and
the two correlated photons are prepared in the state
\begin{equation}
| \Psi \rangle = \int_0^{\infty} {\rm d} \Omega \,
\alpha(\Omega) \int_0^{\Omega} {\rm d}\omega \,
f(\omega) f(\Omega-\omega) \, 
\hat{a}^{\dagger}_{\rm I}(\omega) \,
\hat{a}^{\dagger}_{\rm II}(\Omega-\omega) \, | 0 \rangle,
\label{3.13}
\end{equation}
where $\alpha(\Omega)$ and $f(\omega)$ are the bandwidth 
functions of the laser and down-conversion photons, respecteively,
$f(\omega)$ being centered at $\Omega/2$. From photodetection theory 
it is well known (see, e.g., \cite{vowe}) that the overall
coincidences $R$ can be obtained from the time-integrated
normally ordered intensity correlation function,
\begin{equation}
R = \xi^2 \int {\rm d}t_1 \int {\rm d}t_2 \,
\left\langle
\hat{E}^{(-)} ( t_1 )
\hat{E}^{(-)} ( t_2 )
\hat{E}^{(+)} ( t_1 )
\hat{E}^{(+)} ( t_2 )
\right\rangle\!,
\label{3.13a}
\end{equation}
where $\hat{E}^{(\pm)}(t_1)$ and $\hat{E}^{(\pm)}(t_2)$ are the
fields at the detectors in the two output channels of the beam splitter
($\xi$, detection efficiency). Applying the input--output relations
(\ref{3.10}) and using Eq.~(\ref{3.13}), after some lengthy but
straightforward calculation we find that  
\begin{equation}
R = 2 \pi^2 {\cal N}^4 \int_0^{\infty} {\rm d} \Omega 
\,\alpha^2 (\Omega) \, F(\Omega), 
\label{3.15}
\end{equation}
\begin{equation}
F(\Omega) = 
\int_{0}^{\Omega} \! {\rm d} \omega 
\!\left|f^2 (\omega) f^2 (\Omega \! - \! \omega)\right| 
\omega (\Omega \! - \! \omega)
T^*_{12} (\Omega \! - \! \omega) 
\!\left[T_{12} (\Omega \! - \! \omega)  
\! - \! e^{-2 i \Omega s} e^{4 i\omega s} T_{12} (\omega) \right]\!, 
\quad
\label{3.15a}
\end{equation}

\vspace{.5ex}

\noindent
where $s$ is the translation length of the prism (cf. Fig.~1), and
the abbreviation $ {\cal N}$ $\! = $ $\! \sqrt{\xi \hbar/(4\pi c 
\epsilon_0 {\cal A})} $ has been introduced.

The translation length $s$ $\!=$ $s_0$ that corresponds
to the minumum of $R(s)$ is usually used in order to distinguish between 
superluminal (positive values of $ s_0 $) and subluminal behaviour 
(negative values of $ s_0 $) of the photon passing through the barrier.
In the numerical calculations we have considered H(LH)$^k$ structured 
plates (H, titanium dioxide; L, fused silica) of $\lambda/4$-layers, 
which are of the type described in Ref.~\protect\cite{CKS}. We have
calculated the function $T(\omega)$ applying the algorithm given 
in Ref.~\cite{19}. Due to the lack of reliable data
we have assumed that in the (relevant) frequency interval the
complex refractive indices are approximately independent of
frequency, so that all the dependences on frequency effectively result
from the geometry of the barrier. For the sake of simplicity we have assumed
that the line shape function of the exciting laser, $\alpha(\Omega)$, is 
sufficiently small, so that $F(\Omega)$ $\!\approx$ $\!F(\omega_0)$
in Eq.~(\ref{3.15}), where $\omega_0$ is the centre frequency 
($\omega_0$ $\!=$ $\!5.37 \times 10^{15}{\rm s}^{-1}$). Introducing the 
single-photon pulse shape function  $f(t)$ $\!=$ $\!(2 \pi)^{-1/2} 
\int {\rm d} \omega$ $\!\exp[-i \omega t] f(\omega)$, we have performed 
calculations for both Gaussian pulses $f(t)$ $\!\propto$
$\! \exp [i \omega_0 t / 2 - ( t/t_0 )^2]$
and time-limited non-Gaussian pulses $f(t)$ $\!\propto$
$\! \exp \{i \omega_0 t / 2 - [1- [t/( 2 t_0) ]^2]^{-1} \} $
if  $|t|$ $\!<2$ $\!t_0$ and  $f(t)$ $\!=$ $0$ elsewhere, where
$t_0$ $\!=$ $\!20\,{\rm fs}$ in either case.

The values of $\Delta \tau$ $\! = $ $\!2 s_0/c$ that are shown in
Fig.~2 are valid for both Gaussian and time-limited
non-Gaussian pulse shape functions. The values are positive and indicate
superluminal behaviour of the photon at the barrier, the characteristic
tunneling time being given by $\tau_{\rm t}$ $\!=$ $\!l/c$ $\!-$
$\!\Delta \tau$ ($l$, thickness of the barrier). From the figure we see
that the ``lead'' of the photon, $\Delta \tau$, increases with the
number of layers of the barrier, $N$ $\!=$ $\!2k$ $\!+$ $\!1$, and tends
to a linear function of $N$, the slope of which sensitively depends on the
losses in the barrier. Disregarding the losses, the slope in the 
linear regime is simply given by the inverse velocity of light in vacuum,
which indicates that $\tau_{\rm t}$ is independent of $N$.
The effect of losses is seen to decrease the slope which implies 
that $\tau_{\rm t}$ is increased.

The interval of $N$ in which $\Delta \tau$ (linearly) increases with $N$
must of course be limited by an upper boundary, which may substantially
change with the pulse shape function of the photon at the entrance plane.
For the system under consideration the increase of $\Delta \tau$ with $N$
ends when $N$ $\!\approx $ $\!35$ (lossless case) or 
$N$ $\!\approx $ $\!41$ (lossy case) for the time-limited non-Gaussian 
pulse, whereas for the Gaussian pulse the boundary
value of $N$ is substantially increased.
It should be pointed out that the observed increase of $\Delta \tau$ 
with $N$ is not in contradiction to causality. 
The effect can simply be explained by a shift of the pulse maximum 
towards earlier times owing to pulse reshaping in the barrier, where
the earliest time is given by the time at which the pulse starts from
zero. Since in the case of a Gaussian pulse the pulse 
maximum can be shifted to earlier times than in the case of a 
time-limited pulse, in the former case the upper boundary 
of $N$ is higher than in the latter case.

For sufficiently large $N$ the introduction of the
time $\tau$ (and the traversal time $\tau_{\rm t}$) makes 
little sense. In Fig.~3 the transmittance of the barrier, 
$T_{12}(\omega)$, is plotted for relatively low (11) and high (41) numbers 
of layers. Since the spectral line shape function of the outgoing photon,
$\overline f(\omega)$ $\!\propto$ $\! f(\omega) T_{12} (\omega)$,
sensitively depends on the two competing quantities $f(\omega)$ and
$T_{12}(\omega)$ (Figs.~4, 5), it
can essentially be different from that of the incoming photon
when the number of layers is large enough (compare Fig.~4 with
Fig.~5). The behaviour in the time domain is illustrated
in Figs.~6 and 7 in which the intensity of the
outgoing photon, $\overline{I}(t)$ $\!=$
$\langle \hat{E}^{(-)} \hat{E}^{(+)} \rangle$, is plotted.
In particular, for sufficiently large $N$ the incoming and outgoing 
photons' wavepackets lose all resemblance to each other
(cf. Figs.~5 and 7). In this case the measured 
coincidences are expected to be more or less complicated functions of 
the translation length, the structure of which does not allow one to 
define uniquely a traversal time.

Figures 4 -- 7 refer to the case when the pulse of 
the incoming photon is time-limited. Compared to such a pulse, the 
spectral line shape function $f(\omega)$ of a Gaussian pulse is more 
smoothed and its wings decrease substantially faster. 
Hence, the transformed line shape function
$f(\omega) T_{12} (\omega)$ of a Gaussian pulse
reflects the frequency response of the transmittance of the barrier,
$T_{12} (\omega)$, less sensitively than that of a time-limited
non-Gaussian pulse. This different behaviour explains the above
mentioned difference in the boundary values of $N$.

In Figs.~8 and 9 the coincidences are shown
as a function of the translation length for the chosen time-limited
pulse shape and various numbers of layers. We clearly see that 
when the value of $N$ exceeds an upper boundary value,
then the function $R(s)$ loses the simple structure that can typically
described by a well-defined minimum. It should be noted
that, compared to lossless barries, frequency-selective absorption
shifts the boundary towards higher values.
With increasing $N$ interference fringes are
observed, which correspond to the various possibilities of
(partial) overlapping of the undisturbed and the multi-peaked
outgoing photons' wavepackets at the beam splitter. Clearly,
each minimum introduces its own characteristic time, and a
unique tunneling time can be hardly derived in this way.

\section{Summary and Conclusions}
\label{Sum}
On the basis of a Green function approach to the problem of quantization
of radiation in inhomogeneous, dispersive and absorptive linear dielectrics
we have derived quantum optical input-output relations for optical
fields at multilayer dielectric plates, which can be regarded as 
generalizations of the well-known concepts of unitary transformations
that apply to non-absorbing matter.
Applying the theory to photon tunneling through absorbing barriers,
we have shown that relatively small imaginary parts of the
refractive indices of the layers can already give rise to observable 
effects in two-photon interference experiments as performed recently.

The results reveal that only up to an upper boundary for the number of 
layers the measured coincidences can be used for extracting from them a
characteristic time that may be regarded as traversal time.
The boundary value sensitively depends on the spectral line shape
function of the photon at the barrier and the dependence on
frequency of the transmittance of the barrier, which can be
substantially different for absorbing and non-absorbing barriers.
It is worth noting that for sufficiently large numbers of layers
the photon's wavepacket can be distorted in the barrier to such an extend
that the observed coincidences show a number of interference
fringes which correspond to different time constants.

\section*{Acknowledgement}

One of us (T.G.) is grateful to R.Y. Chiao for valuable comments.

\unitlength1mm
\begin{figure}[htb]
%\begin{picture}(10,40)
\begin{picture}(100,75)
%\put(-20,-60){\psfig{figure=tunnel1fig1.ps,height=6.8in,angle=0}}
\put(-20,-100){\psfig{figure=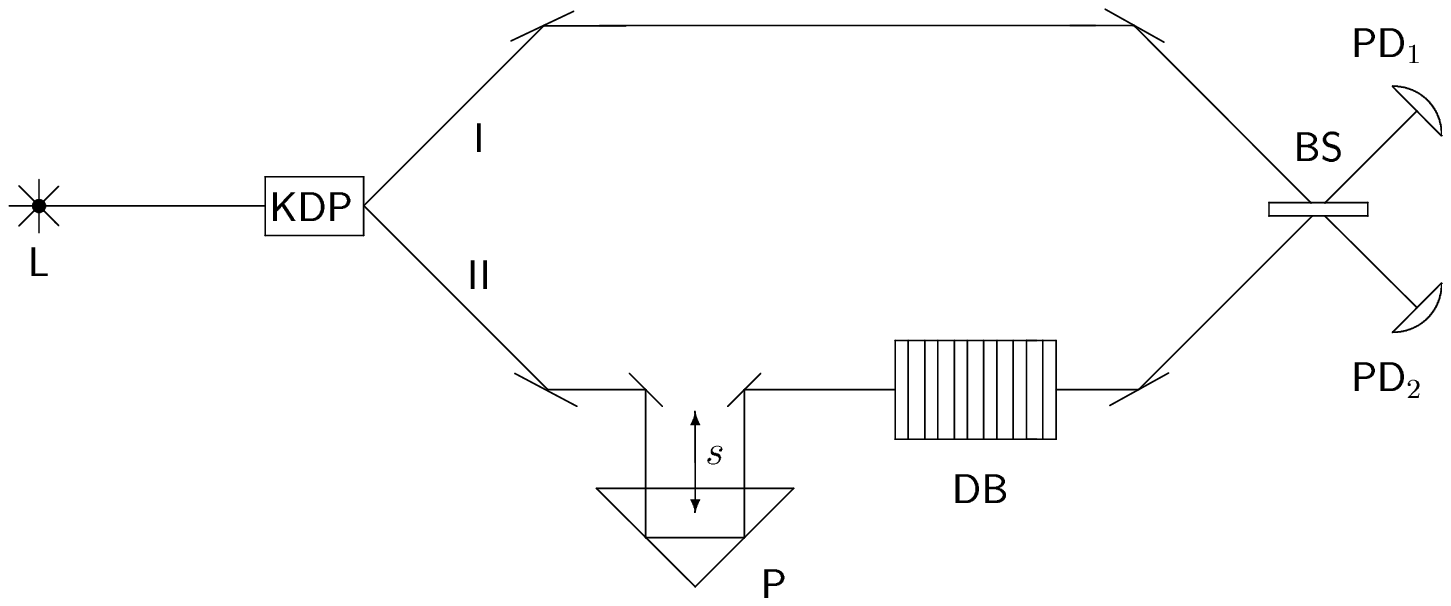,height=10in,angle=0}}
%%%%%%%%%%%%%%%%%%%%%%%%%%%%%%%%%%%%%%%%%%%%%%%%%%%%%%%%%%%%%%%%%%%%%%%%%%%%%%
%%% Erlaeuterung, eingefuegt von T. Gruner 
%%%%%%%%%%%%%%%%%%%%%%%%%%%%%%%%%%%%%%%%%%%%%%%%%%%%%%%%%%%%%%%%%%%%%%%%%%%%%%
% put(x,y) legt die horizontale und vertikale Ausrichtung des Bildes fest
% height legt die Groesse fest (hier als Beispiel auf die Haelfte reduziert)
%%%%%%%%%%%%%%%%%%%%%%%%%%%%%%%%%%%%%%%%%%%%%%%%%%%%%%%%%%%%%%%%%%%%%%%%%%%%%%
\end{picture}
\caption{\sf
Scheme of the two-photon interference experiment \protect\cite{CKS,17b}
for determining photon traversal times through multilayer
dielectric barriers (L, laser; P, prism; DB, dielectric barrier;
BS, beam splitter; PD$_1$, PD$_2$, photodetectors).
\label{fig1}
}
\end{figure}

\begin{figure}[htb]
\begin{picture}(100,120)
\put(-5,-60){\psfig{figure=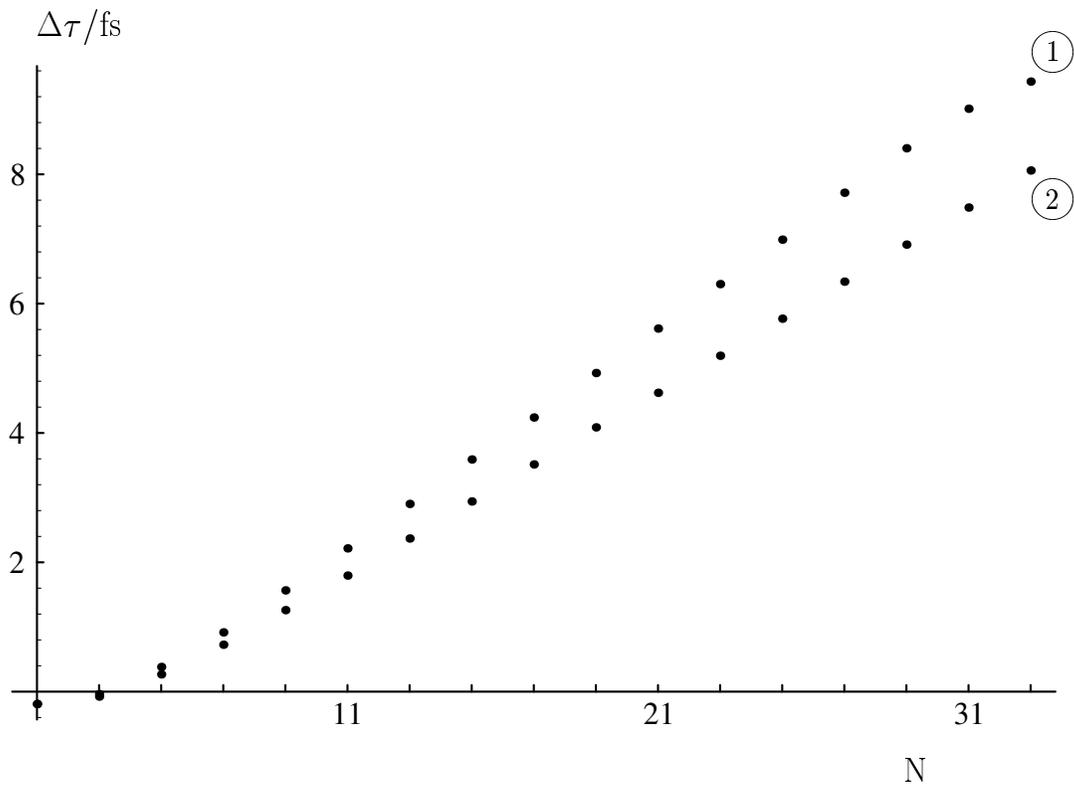,height=9in,angle=0}}
\end{picture}
\caption{\sf
The temporal ``lead" $\Delta \tau $ $\!=$ $\! 2 s_0/c$
that corresponds to the position $s_0$ of the minimum of $R(s)$ is shown
as a function of the number of layers, $ N $ $\!=$ $\!2k$ $\!+$ $\!1$,
for a H(LH)$^k$ structured plate of $\lambda/4$-layers of the
type described in Ref.~\protect\cite{CKS};
curve (1): lossless barrier ($ n_{\rm TiO_2} $ $\!=$ $\! 2.22 $,
$ n_{\rm SiO_2} $ $\!=$ $\! 1.41 $),
curve (2): absorbing barrier ($ n_{\rm TiO_2} $ $\!=$ $\! 2.22 $,
$ n_{\rm SiO_2} $ $\!=$ $\! 1.41 $ $\!+$ $\! 0.0372\,i $
\protect\cite{18}).
\label{fig2}
}
\end{figure}

\begin{figure}[htb]
\begin{picture}(100,120)
\put(-7,-50){\psfig{figure=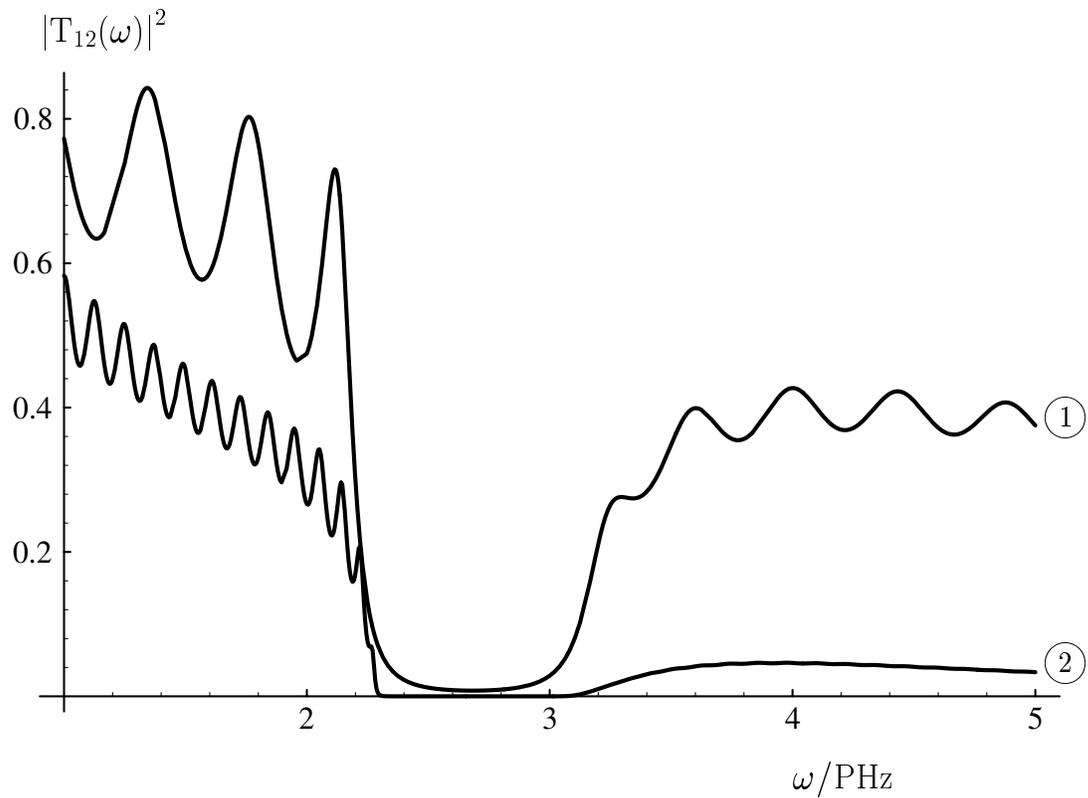,height=9in,angle=0}}
\end{picture}
\caption{\sf
The square of the absolute value of the transmittance of a
multilayer absorbing barrier, $|T_{12}(\omega)|^2$,
is shown for $N$ $\!=$ $\!11$ layers [curve (1)] and $N$ $\!=$ $\!41$ layers
[curve (2)]. The data are the same as in  Fig.~\protect\ref{fig2}.
\label{fig3}
}
\end{figure}

\begin{figure}[htb]
\begin{picture}(100,120)
\put(-7,-50){\psfig{figure=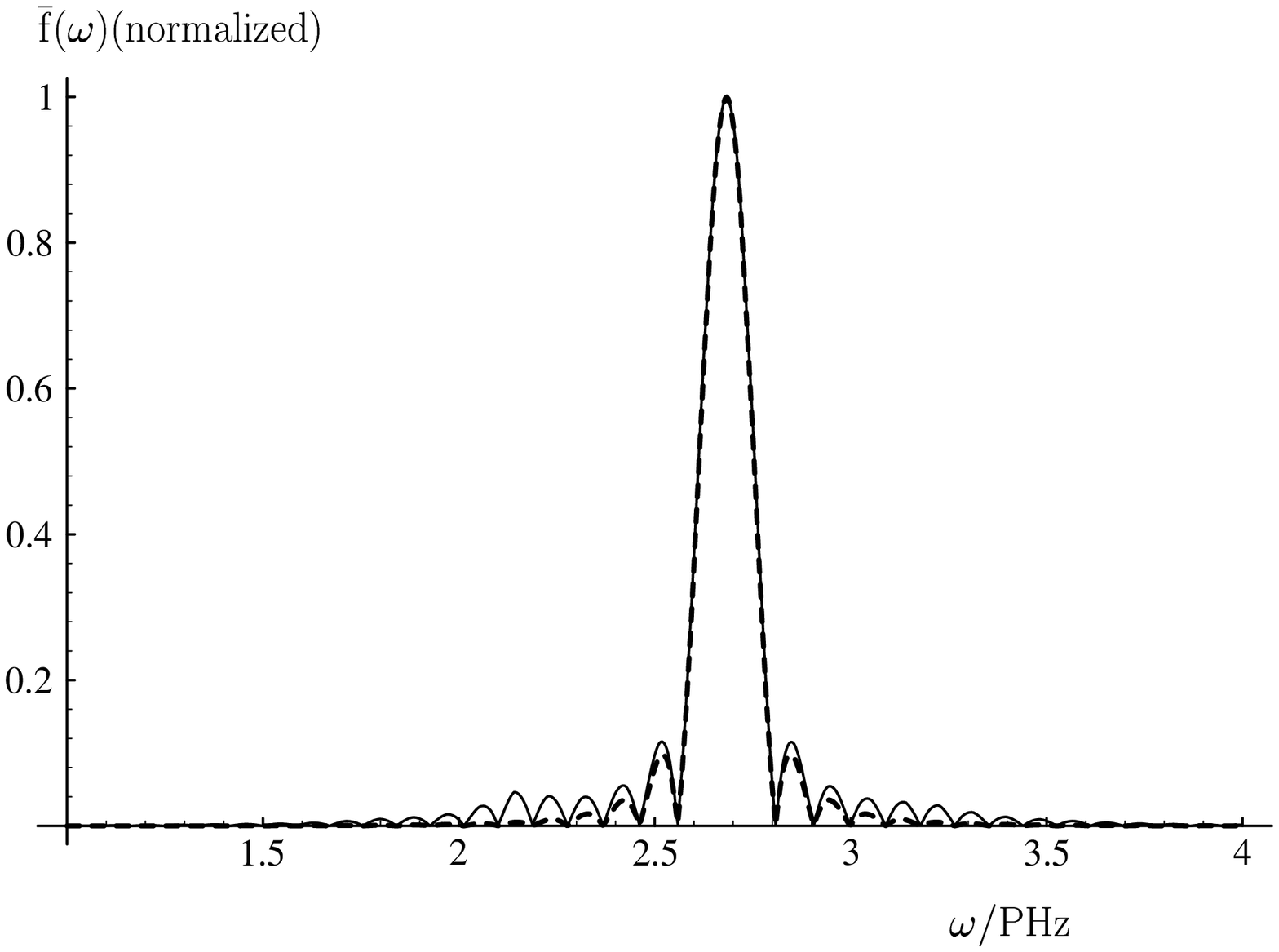,height=9in,angle=0}}
\end{picture}
\caption{\sf
The (normalized) spectral line shape function $\overline f (\omega)$ (full
line) of a photon after having passed through an absorbing barrier consisting
of $N$ $\!=$ $\!11$ layers (data as in  Fig.~\protect\ref{fig2}) is shown.
For comparison, the line shape function of the incoming
pulse that is assumed to be limited in time ($2t_0$ $\!=$ $\!40$\,fs)
is also shown (dashed line).
\label{fig4}
}
\end{figure}

\begin{figure}[htb]

\begin{picture}(100,120)
\put(-7,-50){\psfig{figure=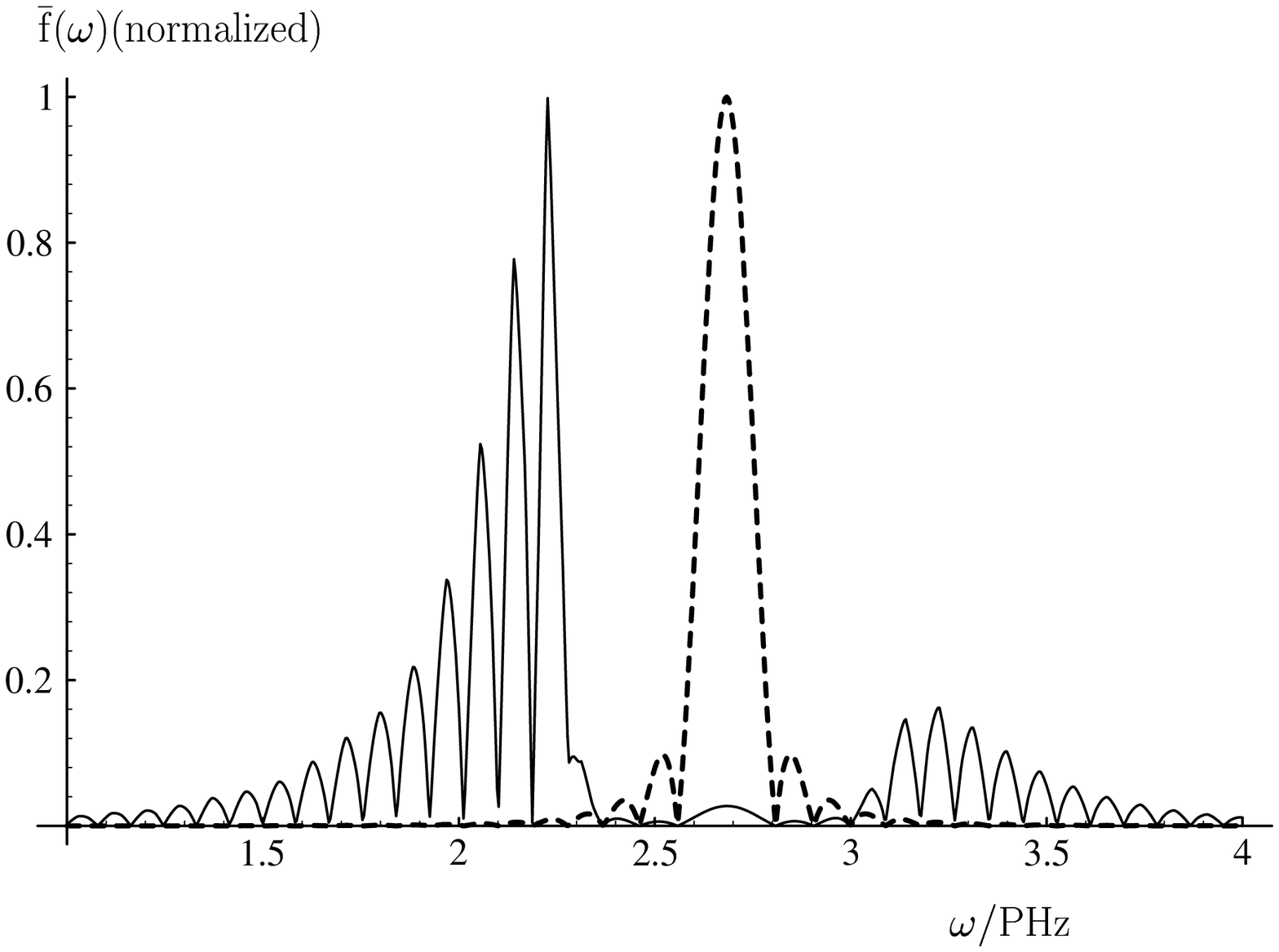,height=9in,angle=0}}
\end{picture}
\caption{\sf
The (normalized) spectral line shape function $\overline f (\omega)$ (full
line) of a photon after having passed through an absorbing barrier consisting
of $N$ $\!=$ $\!41$ layers (data as in  Fig.~\protect\ref{fig2}) is shown.
For comparison, the line shape function of the incoming
pulse that is assumed to be limited in time ($2t_0$ $\!=$ $\!40$\,fs)
is also shown (dashed line).
\label{fig5}
}
\end{figure}

\begin{figure}[htb]
\begin{picture}(100,120)
\put(-7,-50){\psfig{figure=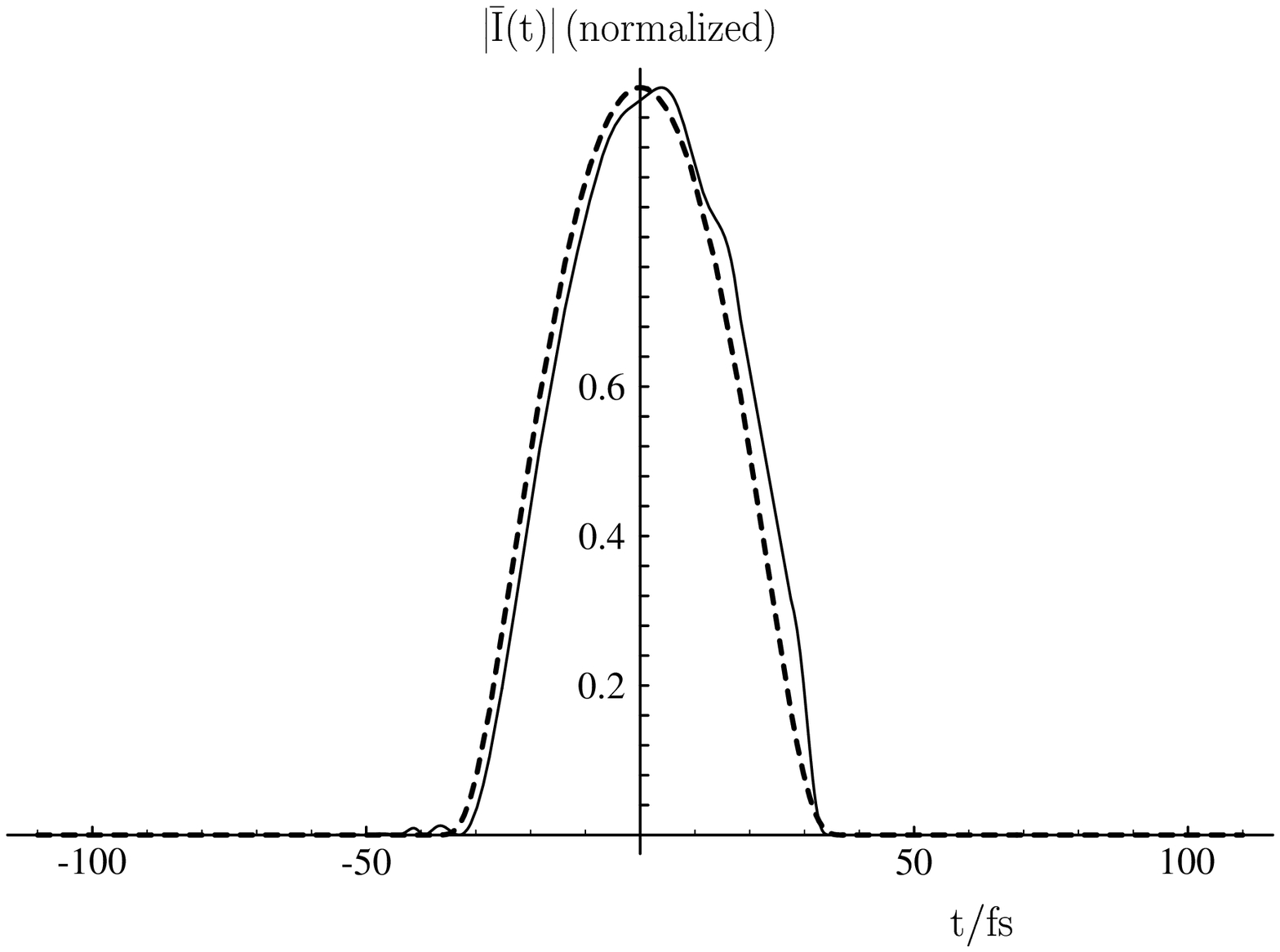,height=9in,angle=0}}
\end{picture}
\caption{\sf
The (normalized) intensity $\overline I(t)$ (full line) of
a photon after having passed through an absorbing barrier consisting of
$N$ $\!=$ $\!11$ layers (data as in  Fig.~\protect\ref{fig2}) is shown.
For comparison, the intensity of the incoming
pulse that is assumed to be limited in time ($2t_0$ $\!=$ $\!40$\,fs)
is also shown (dashed line).
\label{fig6}
}
\end{figure}

\begin{figure}[htb]
\begin{picture}(100,120)
\put(-7,-50){\psfig{figure=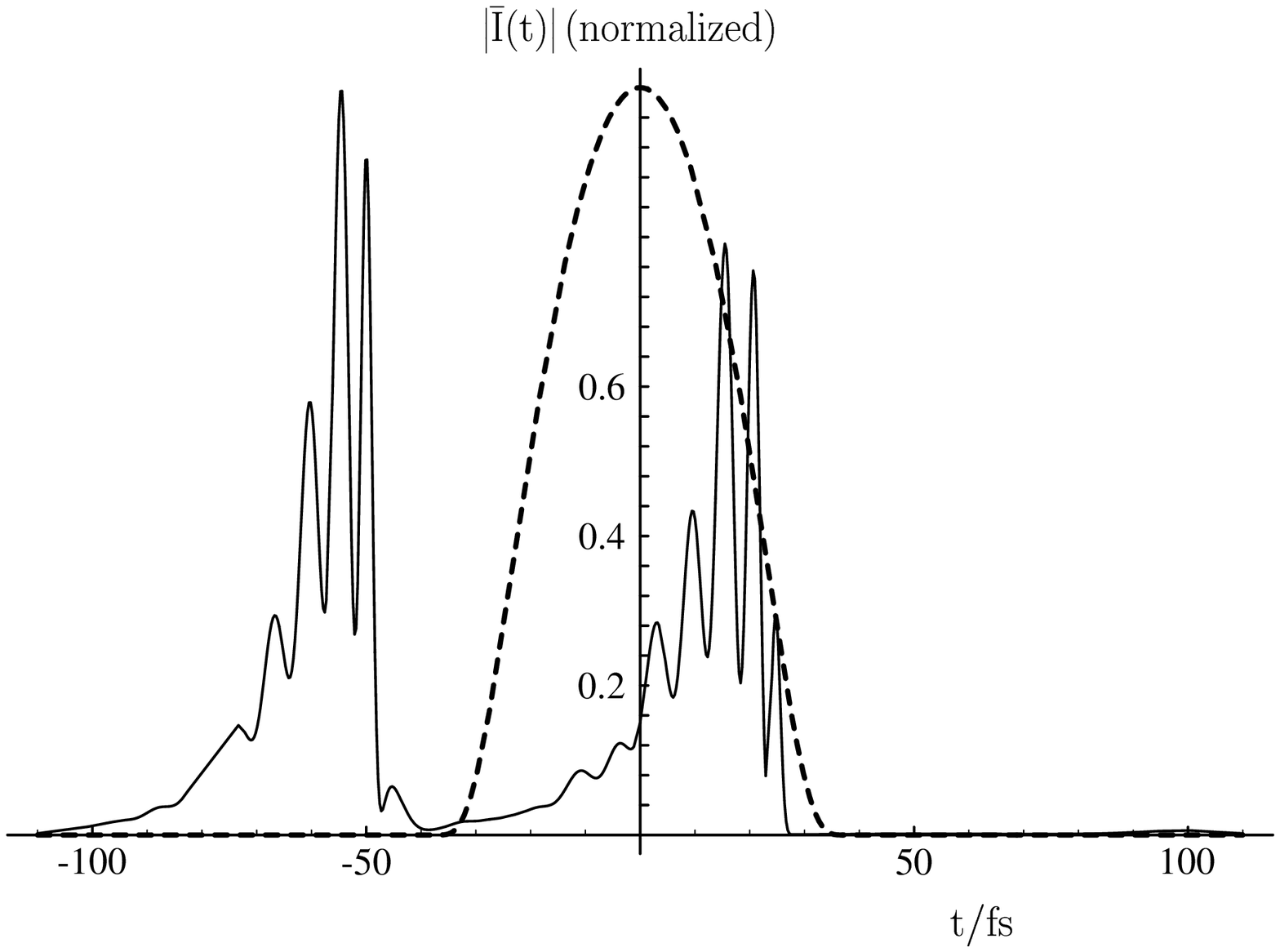,height=9in,angle=0}}
\end{picture}
\caption{\sf
The (normalized) intensity $\overline I(t)$ (full line) of
a photon after having passed through an absorbing barrier consisting of
$N$ $\!=$ $\!41$ layers (data as in  Fig.~\protect\ref{fig2}) is shown.
For comparison, the intensity of the incoming
pulse that is assumed to be limited in time ($2t_0$ $\!=$ $\!40$\,fs)
is also shown (dashed line).
\label{fig7}
}
\end{figure}

\begin{figure}[htb]
\begin{picture}(100,120)
\put(-7,-50){\psfig{figure=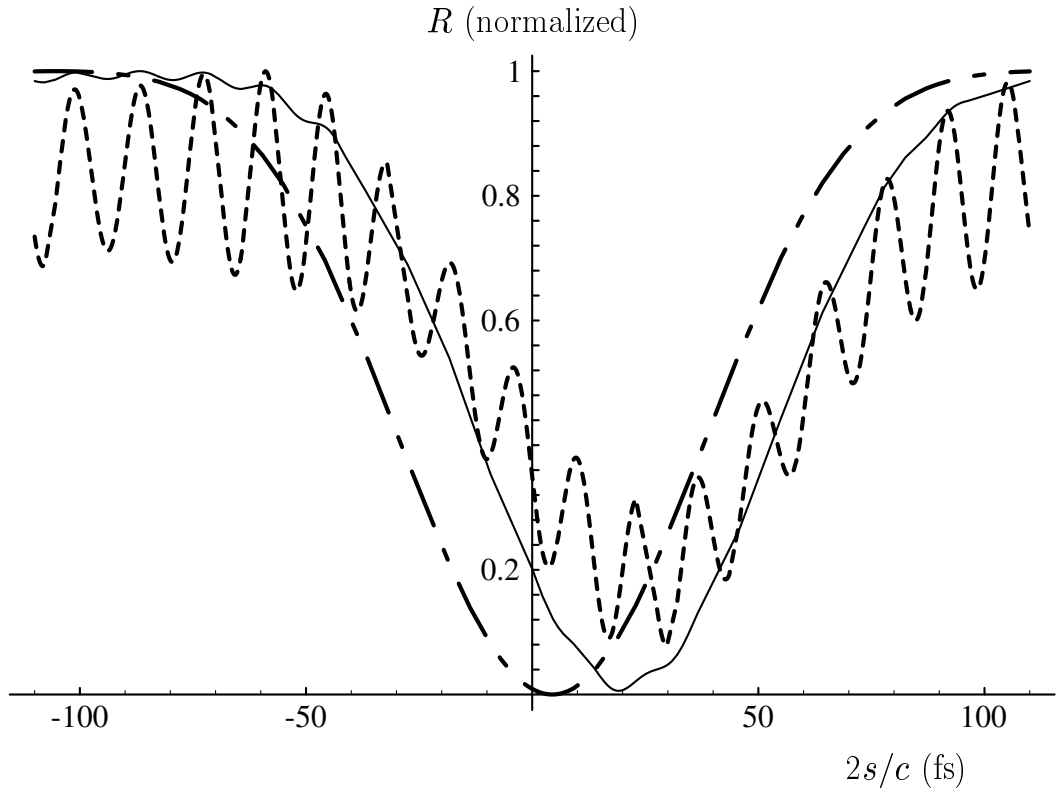,height=9in,angle=0}}
\end{picture}
\caption{\sf
The (normalized) coincidences $R(s)$ are shown in dependence on the
translation length $s$ for a time-limited pulse of the
incoming photon ($2t_0$ $\!=$ $\!40$\,fs) and
various numbers of the layers of a lossless barrier:
$N$ $\!=$ $\!11$ (dotted-dashed line),
$N$ $\!=$ $\!35$ (full line),
$N$ $\!=$ $\!41$ (dashed line).
The data of the lossless barrier are the same as in
Fig.~\protect\ref{fig2}.
\label{fig8}
}
\end{figure}

\begin{figure}[htb]
\begin{picture}(100,120)
\put(-7,-50){\psfig{figure=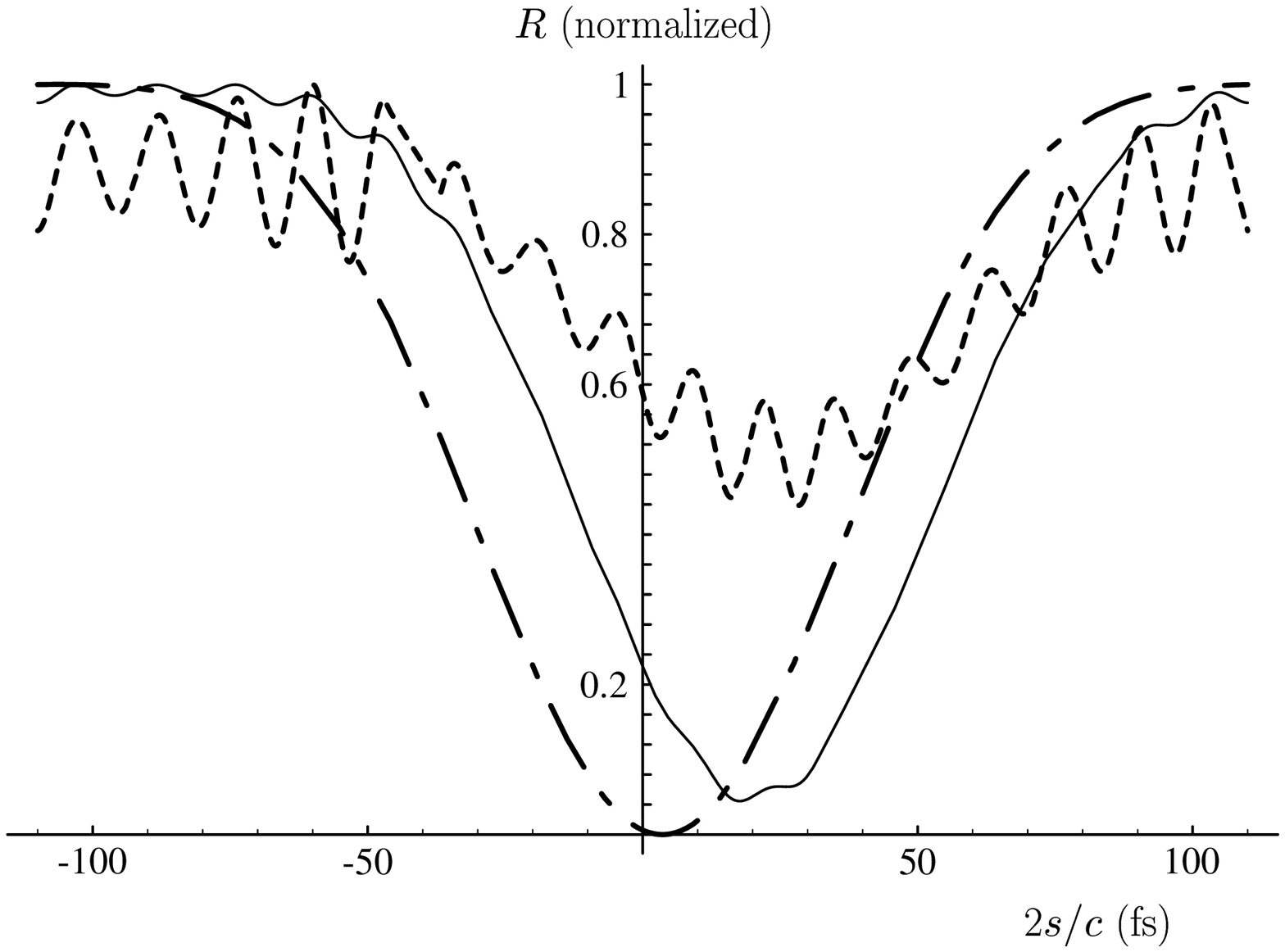,height=9in,angle=0}}
\end{picture}
\caption{\sf
The (normalized) coincidences $R(s)$ are shown in dependence on the
translation length $s$ for a time-limited pulse of the
incoming photon ($2t_0$ $\!=$ $\!40$\,fs) and
various numbers of the layers of an absorbing barrier:
$N$ $\!=$ $\!11$ (dotted-dashed line),
$N$ $\!=$ $\!41$ (full line),
$N$ $\!=$ $\!49$ (dashed line).
The data of the absorbing barrier are the same as in
Fig.~\protect\ref{fig2}.
\label{fig9}
}
\end{figure}

\end{sf}

\begin{thebibliography}{99}

\bibitem{CKS} 
R.Y. Chiao, P.G. Kwiat and A.M. Steinberg, 
Quant. Semiclass. Opt. 7 (1995) 259.

\bibitem{17b}
A.M. Steinberg and R.Y. Chiao, 
Phys. Rev. A 51 (1995) 3525.

\bibitem{1a} 
Ch. Spielmann, R. Szip\"ocs, A. Stingl and F. Krausz, 
Phys. Rev. Lett. 73 (1994) 2308.

\bibitem{4n} 
Y. Japha and G. Kurizki, 
Phys. Rev. A bf 53 (1996) 586.
 
\bibitem{5n} 
A.M. Steinberg, 
Phys. Rev. A 52 (1995) 32.

\bibitem{4}
L. Kn\"{o}ll, W. Vogel and D.--G. Welsch,
Phys. Rev. A 42 (1990) 503;
J. Opt. Soc. Am. B 3 (1986) 1315.

\bibitem{1}
L. Kn\"{o}ll, W. Vogel and D.-G. Welsch,
Phys. Rev. A 36 (1987) 3803.

\bibitem{2}
R.J. Glauber and M. Lewenstein,
Phys. Rev. A 43 (1991) 467.

\bibitem{3}
H. Khosravi and R. Loudon,
Proc. R. Soc. Lond. Ser. A 433 (1991) 337;
{\em ibid.} 436 (1992) 373.

\bibitem{8}
M. Fleischhauer and M. Schubert,
J. Mod. Opt. 38 (1991) 677.

\bibitem{8a}
G.S. Agarwal, 
Phys. Rev. A 11 (1975) 230.

\bibitem{9}
B. Huttner and S.M. Barnett,
Europhys. Lett. 18 (1992) 487;
Phys. Rev. A 46 (1992) 4306.

\bibitem{10}
L. Kn\"{o}ll and U. Leonhardt,
J. Mod. Opt. 39 (1992) 1253.

\bibitem{10a}
D. Kupiszewska, 
Phys. Rev. A 46 (1992) 2286.

\bibitem{11}
S.-T. Ho and P. Kumar,
J. Opt. Soc. Am. B 10 (1993) 1620.

\bibitem{11a}
J.R. Jeffers, N. Imoto and R. Loudon,
Phys. Rev. A 47 (1993) 3346.

\bibitem{12}
T. Gruner and D.-G. Welsch,
Phys. Rev. A 51 (1995) 3246.

\bibitem{13}
S.M. Barnett, R. Matloob and R. Loudon,
J. Mod. Opt. 42 (1995) 1165.

\bibitem{13a}
R. Matloob, R. Loudon, S.M. Barnett and J. Jeffers,
Phys. Rev. A 52 (1995) 4823.

\bibitem{14a} 
T. Gruner and D.-G. Welsch, 
Proceedings of the
Third Workshop on Quantum Field Theory under the Influence of
External Conditions, Leipzig, 1995 (ed. M. Bordag, 
B.G. Teubner Verlagsgesellschaft, Stuttgart, Leipzig, 1996).

\bibitem{14} 
T. Gruner and D.-G. Welsch,
Phys. Rev A 53 (1996) 1818.


\bibitem{BW}
M. Born and E. Wolf,
Principles of Optics (Pergamon Press, London, 1959).

\bibitem{19}
T. Gruner and D.-G. Welsch, 
quant-ph/9511041;
Phys. Rev. A, to be published.

\bibitem{18}
H.R. Philipp, EMIS Datareview, Aug. 1987.

\bibitem{vowe}
W. Vogel and D.-G. Welsch,
Lectures on Quantum Optics (Akademie Verlag, Berlin 1994).

\end{thebibliography}
\end{document}